\documentclass{article}

\usepackage{amsmath, amssymb}
\usepackage{graphicx}
\usepackage[mathscr]{eucal}

\newcommand{\rf}[1]{(\ref{#1})}
\newcommand{\ba}{\begin{array}}
\newcommand{\ea}{\end{array}}
\newcommand{\dis}{\displaystyle}

\newcommand{\bra}[1]{\langle\,#1\,|}
\newcommand{\ket}[1]{|\,#1\,\rangle}
\newcommand{\p}{\partial}
\newcommand{\ud}{\mathrm{d}}
\newcommand{\pathD}{\!\mathscr{D}}

\newcommand{\f}{\varphi}

\newcommand{\x}{{\bf x}}
\newcommand{\y}{{\bf y}}
\newcommand{\z}{{\bf z}}

\newcommand{\X}{\mathbf{X}}

\numberwithin{equation}{section}

\title{The vacuum state functional of interacting string field theory}
\author{A. Ilderton\\ \\Centre for Particle Theory, University of Durham, Durham DH1 3LE, UK\\ \texttt{a.b.ilderton@dur.ac.uk}}
\date{}
\begin{document}
\maketitle
\abstract{We show that the vacuum state functional for both open and closed string field theory can be reconstructed from the vacuum expectation values it must generate. The method also applies to quantum field theory and as an application we give a diagrammatic description of the equivalence between Shr\"odinger and covariant representations of field theory.}

\section{Introduction and review}
The Euclidean vacuum functional of quantum field theory \cite{Abers} and first quantised string theory \cite{Birm} can be constructed via a large time functional integral. The Lorentzian continuation in field theory is
\begin{equation}
\Psi_0[\phi] = \int\pathD\f\,\, \exp\bigg(\frac{i}{\hbar}\int\limits_{-\infty}^0\!\ud x^0\,\, L(\f,\dot\f)\bigg)\bigg|^{\f=\phi\text{ at time $0$}}_.
\end{equation}
The boundary condition in the infinite past is that the field must be regular. By shifting the integration variable by a piece proportional to the Heaviside function the field dependence is moved out of the boundary conditions and into the action, giving
\begin{equation}\label{v-def}  
  \Psi_0[\phi] = \int\pathD\f\,\, \exp\bigg(\frac{i}{\hbar}\int\limits_{-\infty}^0\!\ud x^0\,\, L(\f,\dot\f) + \frac{i}{\hbar}\int\!\ud^D\x\,\,\dot\f(\x,0)\phi(\x)\bigg)\bigg|^{\f=0\text{ at time $0$}}.
\end{equation}
Standard field theory results then imply that the logarithm of the vacuum functional is the sum of connected Feynman diagrams constructed from a propagator $G_d$ which obeys Dirichlet boundary conditions on the hypersurface $x^0=0$ with vertices integrated over the interval $x^0<0$. All external legs (those attached to $\phi$) end on the boundary with a time derivative which results from $\phi$ being coupled to $\dot\f$.

Our aim is to construct the vacuum state functional for string field theory, following the interest generated by the Sen conjectures \cite{Sen}. The string field interaction will be non-local in our chosen time co-ordinate, but the above description of the vacuum fails if the interaction becomes non-local. This is because we would attempt to describe the functional by a sum over field histories on the half space $x^0<0$, but through a non-local interaction the field can couple to itself at arbitrary times. 

Following this introduction we will describe an alternative method of constructing the vacuum functional which applies to both local and non-locally interacting field theories, and then generalise to string field theory. The essential ingredient is the set of gluing properties. For the scalar field propagator these are
\begin{equation}\label{prop-cases}
\int\!\ud^D \y \,\, G_0(\x_2,t_2;\y,t)\left(-2\frac{\p}{\p t}\right)G_0(\y,t;\x_1,t_1)=
\left\{ \ba{cr}
{\dis  iG_0(\x_2,t_2;\x_1,t_1)} & t_2> t>t_1
\\
{\dis  -iG_0(\x_2,t_2;\x_1,t_1)} & t_1>t>t_2
\\
{\dis  iG_I(\x_2,t_2;\x_1,t_1)} & t>t_1,t_2
\\
{\dis  -iG_I(\x_2,t_2;\x_1,t_1)} & t<t_1,t_2
\ea
\right.
\end{equation}
These rules apply to both the open and closed string field propagator \cite{Polch} \cite{Ord} when time is associated with the hypersurfaces $X^0(\sigma)=\text{ constant}$. The point $\x^i$ is replaced by the 25 dimensional spacetime curve $\X^i(\sigma)$ along with the non-vanishing ghost components on the Dirichlet sections of the propagator. In this paper $\X$ will be shorthand for this whole set to simplify this notation. For a proof of the gluing rules and a discussion of the ghost sector see \cite{us2}.

We will construct the vacuum wave functional using the vacuum expectation values it must generate. This will apply to both open and closed string field theories with a cubic interaction, though the explicit form of this interaction is arbitrary. Our construction is perturbative, we give the first few terms of the functional explicitly, but the procedure used generalises to all orders in perturbation theory. In the final section of this paper we will return to local field theory and use our gluing rules to give a diagrammatic demonstration of the equivalence of Schr\"odinger and covariant field theory pictures in field theory.

\section{The string field vacuum state}
The vacuum state functional is the generator of vacuum expectation values. For example, the open string three field expectation value would be represented by, to leading order in the coupling, the diagram

\begin{equation}
  \includegraphics[height=2.2cm]{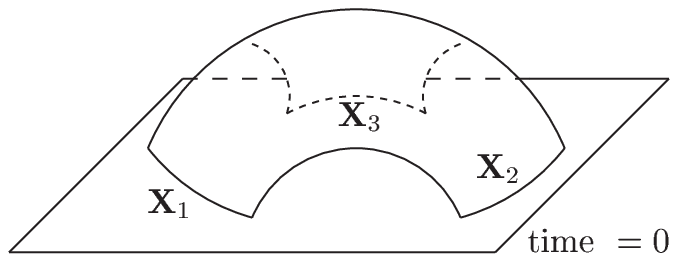}
\end{equation}
Such expectation values can be used to reconstruct the vacuum state functional. We will demonstrate this for quantum field theory and then generalise to strings. Since the string field interaction is in various guises cubic we will consider a non-local scalar $\phi^3$ theory with interaction Hamiltonian density
\begin{equation}
  \frac{\lambda}{3!} \int\!\prod\limits_{j=1}^3\ud^D\x_j\ud t_j\,\, W(t, \x;\x_1,t_1, \x_2,t_2,\x_3,t_3)\phi(\x_1,t_1)\phi(\x_2,t_2)\phi(\x_3,t_3)
\end{equation}
written in terms of some kernel $W$. Replacing this kernel with a product of delta functions $\delta(t-t_i)$ recovers the local $\phi^3$ Hamiltonian density. The technique we are about to describe applies to both local and non-local interactions, as demonstrated in \cite{us2} for local $\phi^4$ theory.

To begin, consider the free field vacuum functional, which must yield the equal time propagator as a vacuum expectation value,
\begin{equation*}
  \hbar G(\x,0;\y,0) = \bra{0}\phi(\x,0)\phi(\y,0)\ket{0} = \int\pathD\phi\,\, \phi(\x)\phi(\y)\Psi_0^2[\phi]
\end{equation*}
which implies that the free field vacuum is the exponent of the inverse of the equal time propagator. This inverse is
\begin{equation}\label{inverse}
  G_0(\x,0;\y_0)^{-1} = 4\frac{\p^2}{\p t\p t'}G_0(\x,t;\y_t')\bigg|_{t=t'=0}
\end{equation}
as we proved in \cite{us2}, so the free field vacuum is
\begin{equation}\label{2.4}
  \Psi_0^\text{free}[\phi] = \exp\bigg(-\frac{1}{4\hbar} \int\!\ud^D(\x,\y)\phi(\x)G_0(\x,\overset{\bullet}{0};\y,\overset{\bullet}{0})\phi(\y)\bigg),
\end{equation}
as is easily checked. The bullet is a time derivative with factor $-2$, as in (\ref{prop-cases}). Now introduce the cubic interaction. Our arguments are based in the Schr\"odinger representation, and treats all spatial arguments in the same way as covariant methods. Therefore we may choose the interaction to be local or non-local in space without affecting the results, as such we will write the Hamiltonian as
\begin{equation}
  \frac{\lambda}{3!} \int\!\prod\limits_{j=1}^3\ud t_j\,\, W(t;t_1, t_2,t_3)\phi(t_1)\phi(t_2)\phi(t_3)
\end{equation}
to keep notation compact, suppressing the spatial dependencies whenever possible. Accordingly we will abbreviate our representation of the free propagator to
\begin{equation}\begin{split}
  G_\x(0;t_j) &:= G_0(\x,0;\x_j,t_j), \\
  \phi_\x G_\x(0;t_j) &:= \int\!\ud^D\x\,\,\phi(\x)G_0(\x,0;\x_j,t_j).
\end{split} \end{equation}
To clarify our arguments we also assume that the interaction kernel is symmetric in its indices. The following applies when this is not the case, the only difference being the appearance of sums of terms corresponding to inequivalent ways of attaching propagators. We will now describe how to identify higher order terms in the vacuum functional by expanding in orders of $\lambda$ and $\hbar$ with a recursive process using vacuum expectation values.

We begin with the lowest order expectation value, which is the tree level three field expectation value of order $\lambda\hbar^2$. A covariant field theory calculation implies that this is
\begin{equation}\begin{split}
  \bra{0}\phi(\x,0)\phi(\y,0)\phi(\z,0)\ket{0} = -i\lambda\hbar^2\int\limits_{-\infty}^\infty\! \ud t\!\int\limits_{-\infty}^\infty\!&\prod\limits_{j=1}^3 \ud t_j \,\,\,W(t;t_1,t_2,t_3) \\
  \times &G_\x(0;t_1)G_\y(0;t_2)G_\z(0;t_3)
\end{split}\end{equation}
Alternatively, we can abandon the explicit kernel $W$ and think of this as some Feynman diagram which ties together three legs ending at $x^0=0$, in some way (the usual Feynman  diagram showing three propagators meeting at a point vertex is not suitable):
\begin{equation}
  \includegraphics[height=1.6cm]{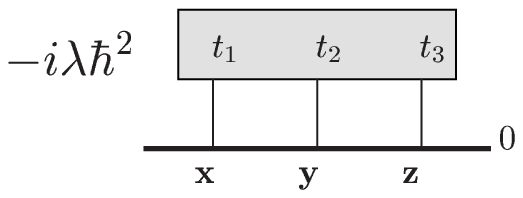}
\end{equation}
Introducing an unknown diagram $\Gamma$ into the vacuum,
\begin{equation}
  \includegraphics[height=1.4cm]{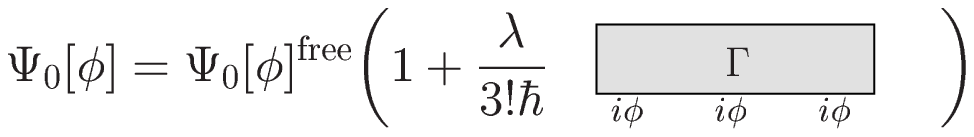},
\end{equation}
we require the expectation value to be given by
\begin{equation}
  \bra{0}\phi(\x,0)\phi(\y,0)\phi(\z,0)\ket{0} = \int\pathD\phi\,\, \phi(\x)\phi(\y)\phi(\z)\Psi_0^2[\phi].
\end{equation}
Equivalently this may be expressed as

\begin{equation}
    \includegraphics[height=1.8cm]{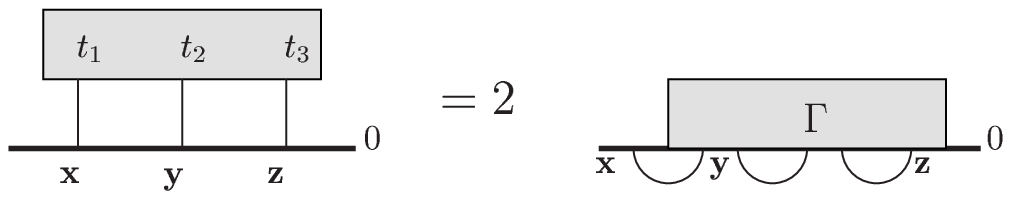}
\end{equation}
where the fields in $\Gamma$ have been contracted with the fields at external points using the inverse of the quadratic piece of the vacuum functional squared, which is the equal time propagator (the arcs on the right hand side). The factor of $2$ appears because of this squaring.

To identify $\Gamma$ we invert these using \rf{inverse}. The right hand side of the above becomes the unknown kernel. The left hand side has external legs attached to the inverse of the equal time propagator. This produces propagators with a time derivative at the boundary $t=0$, and is independent of the sign of the time $t_i$ at the other end of the propagator (see \cite{us2} for details). The result is

\begin{equation}
  \includegraphics[height=1.8cm]{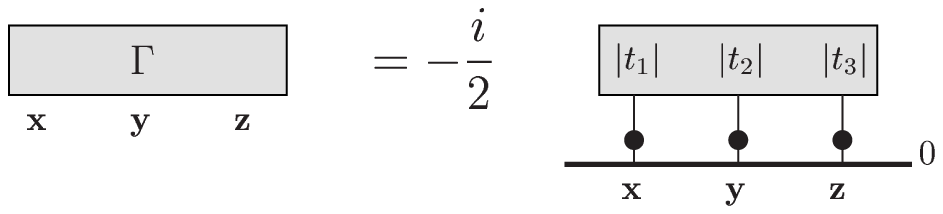}.
\end{equation}
Note the modulus signs on the right hand side, illustrating the independence of the propagator endpoint on the sign of the time. In terms of the interaction kernel $W$ the vacuum state functional is 
\begin{equation}
  \Psi_0[\phi]=\Psi_0^\text{free}[\phi] \bigg(1- \frac{\lambda}{2.3!\hbar}\int\limits_{-\infty}^\infty\!\ud t\prod\limits_{j=1}^3 \ud t_j\,\,\, W(t;t_1,t_2,t_3) \prod\limits_{j=1}^3 \phi_\x\cdot G_{\x}(\overset{\bullet}{0};|t_j|)\bigg).
\end{equation}
The tree level $n$--field term in the vacuum functional can be constructed similarly from the tree level $n$--field expectation value. Let us now describe a one loop term. We construct the vacuum functional tadpole using
\begin{equation}
\langle\,0\,|\phi(\mathbf{x},0)|\,0\,\rangle = -i\frac{\lambda\hbar}{2!}\int\!\mathrm{d}t\prod\limits_{j=1}^3\mathrm{d}t_j\,\,W(t;t_1,t_2,t_3)G_\mathbf{x}(0;t_1)G(t_2;t_3).
\end{equation}
This is the generalisation of the tadpole
\begin{equation}
  \includegraphics[height=1.6cm]{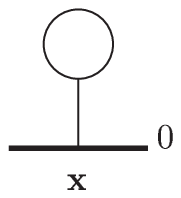}
\end{equation}
in local cubic theory. We illustrate it by
\begin{equation}
  \includegraphics[height=1.6cm]{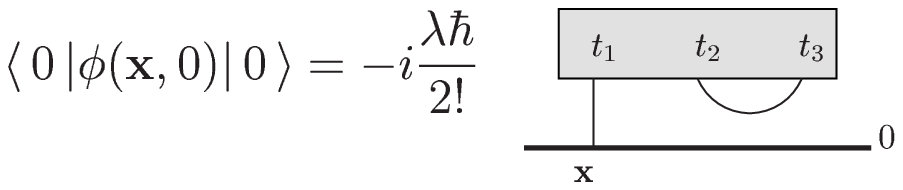}.
\end{equation}
We have explicitly written in the factor of $1/2!$ normally implicit in a Feynman loop. The diagram is of order $\hbar\lambda$, and receives contributions from the single field one loop term in the vacuum state and from the three field tree level piece we constructed above.

We include a new unknown in the vacuum,
\begin{equation}
  \includegraphics[height=1.8cm]{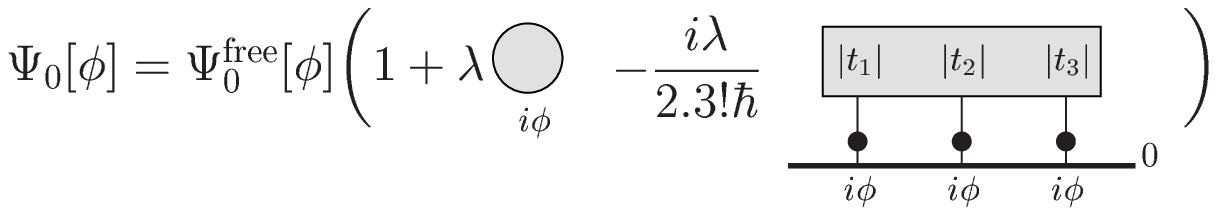},
\end{equation}
and calculating the expectation value implies

\begin{equation}
  \includegraphics[height=1.8cm]{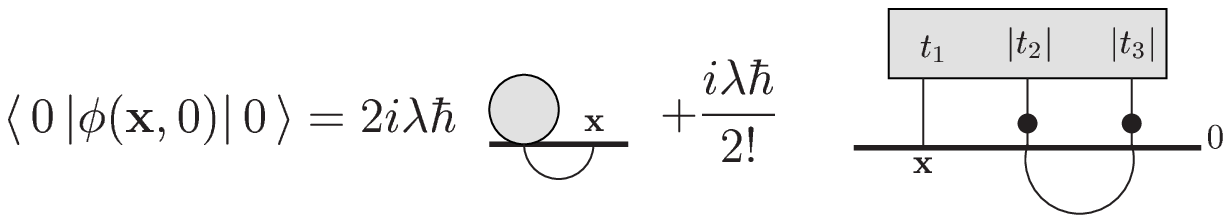}.
\end{equation}
Inverting the equal time propagators as we did before identifies the new kernel,
\begin{equation}
  \includegraphics[height=1.6cm]{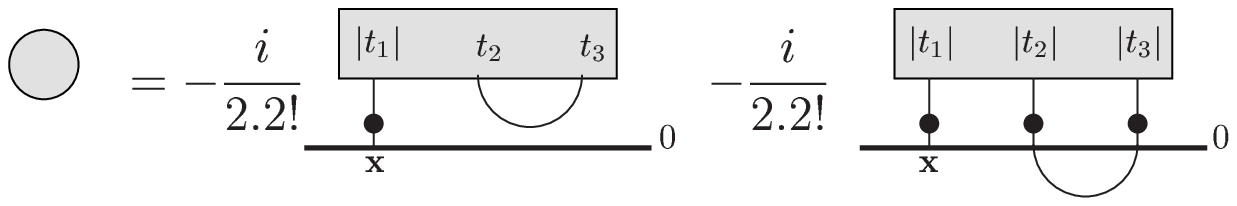}.
\end{equation}
Explicitly,
\begin{equation}\begin{split}
  \Psi_0[\phi]=\Psi_0^\text{free}[\phi] \bigg(1&+\frac{\lambda}{2.2!}\int\!\ud t\prod\limits_{j=1}^3 \ud t_j\,\,W(t;t_1,t_2,t_3)\phi_\x G_\x(\overset{\bullet}{0};|t_1|)G(t_2;t_3) \\
 +\frac{\lambda}{2.2!}\int\!\ud t\prod\limits_{j=1}^3 \ud t_j\,\,&W(t;t_1,t_2,t_3)\phi_\x G_\x(\overset{\bullet}{0};|t_1|) G(\overset{\bullet}{0};|t_2|)G(0;0)G(\overset{\bullet}{0};|t_3|) \\
  &- \frac{\lambda}{2.3!\hbar}\int\limits_{-\infty}^\infty\!\ud t\prod\limits_{j=1}^3 \ud t_j\,\,\, W(t;t_1,t_2,t_3) \prod\limits_{k=1}^3 \phi_\x G_\x(\overset{\bullet}{0};|t_k|)\bigg)
\end{split}\end{equation}
This is very similar to the result found for the local theories in \cite{us2}. Now we repeat the above arguments to construct the string field vacuum. The free field vacuum is (suppressing the $\hbar$ dependence)
\begin{equation}
  \includegraphics[height=1.3cm]{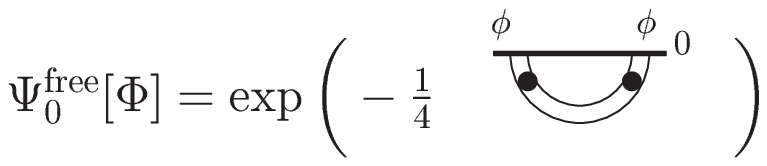},
\end{equation}
just as in (\ref{2.4}). The double represents the open or closed string field propagator. We propose the lowest order expansion
\begin{equation}\label{Gamma}
  \includegraphics[height=2.2cm]{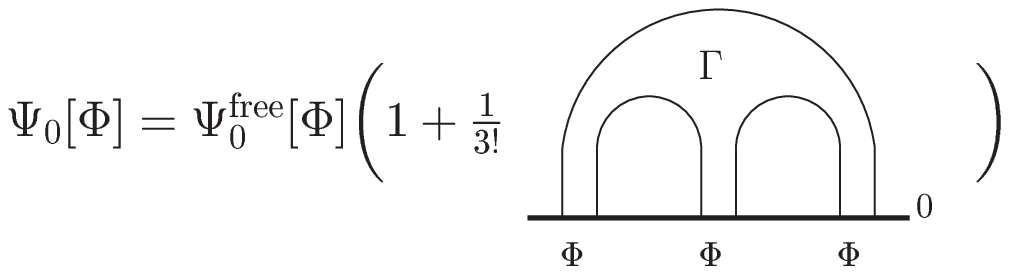}.
\end{equation}
Trying to recover the three field vacuum expectation value using the vacuum functional as we did before, we find

\begin{equation}
  \includegraphics[height=2cm]{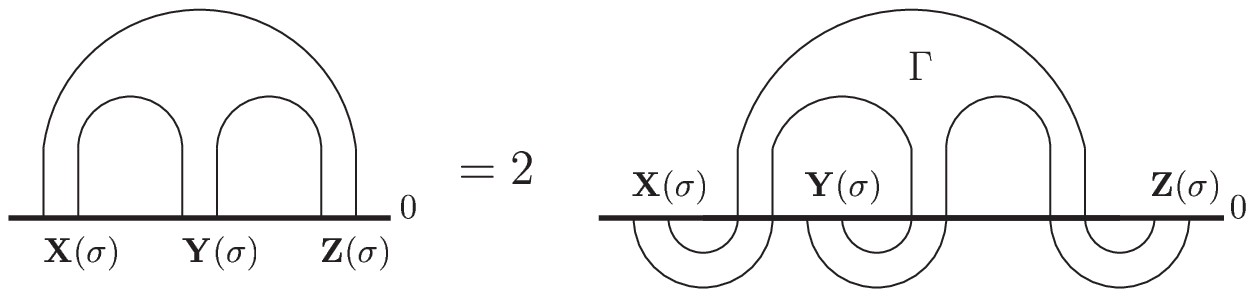}
\end{equation}
The diagram on the left is the three field vacuum expectation value. As before we invert the equal time propagators on the right hand side of the above, and the left hand side becomes the kernel we are trying to identify. The first order cubic term in the vacuum state functional is therefore
\begin{equation}\label{odd}
  \includegraphics[height=2.2cm]{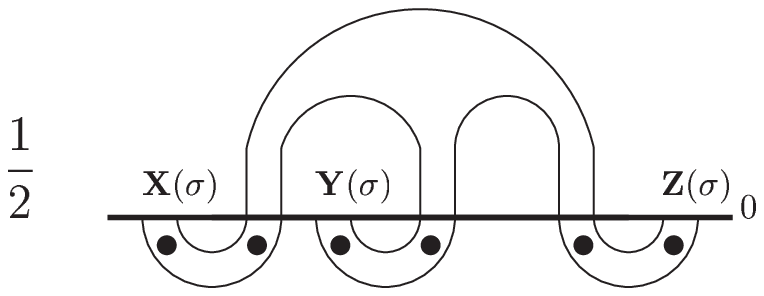}
\end{equation}
Our gluing rules give us the result of joining propagators when each end lies on a constant time surface, they do not apply when one end lies on a time $X^0(\sigma)$ for non-trivial sigma dependence, nor how to attach more general surfaces together. For this we need Carlip's method, which sheds some light on what the above kernel is.

In \cite{Carlip} it was shown that to correctly sew the moduli spaces of two worldsheets the inverse propagator (first quantised Hamiltonian) should be attached to one of the boundaries being sewn. This removes a divergent integral over a redundant length parameter when we integrate over all shared field arguments on the boundaries being sewn.

Let $T\big[t_1',t_2',t_3'\big]$ is the three string amplitude with external legs ending at curves $\X_i(\sigma)$ at times $X^0 = t_i'$ for $i=1\ldots3$. Using the simple identity
\begin{equation}
  f(0) = \int\!\ud s\,\, f(s) \delta(s) = \int\!\ud s\ud q\,\, f(s)G^{-1}(s,q)G(q,0).
\end{equation}
we can write the three field expectation value as
\begin{equation}
  T[0,0,0] = \int\!\prod\limits_{j=1}^3\ud t_j'\ud t_j\,\, T\big[t_1',t_2',t_3'\big]\prod\limits_{i=1}^3G^{-1}(t_i',t_i)G(t_i,0)
\end{equation}
Now when we attach the inverse of the equal time propagator the gluing rules imply that the vacuum functional is
\begin{equation}
  \Psi_0[\Phi] = \Psi_0^\text{free}[\Phi]\bigg(1- \frac{i}{2.3!}\int\!\prod\limits_{j=1}^3\ud t_j'\ud t_j \,\,T\big[t_1',t_2',t_3'\big]\prod\limits_{i=1}^3G^{-1}(t_i',t_i)\Phi_\X\cdot G_\X(\overset{\bullet}{0};|t_k|)\bigg).
\end{equation}
In our case the inverse propagator is that part of the first quantised string theory Hamiltonian which depends on $\X$ and $t$ on the boundary of our worldsheet. We can represent the vacuum functional as

\begin{equation}
  \includegraphics[height=2.8cm]{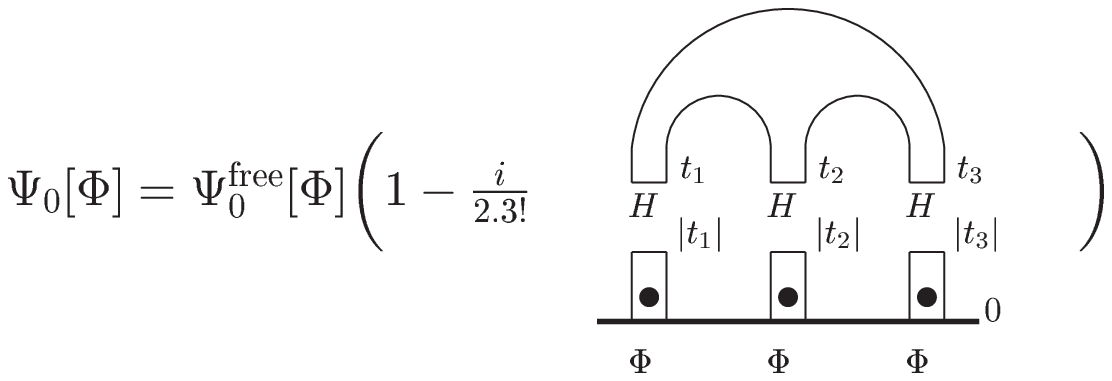}
\end{equation}
where we have written $H$ in place of the inverse propagators for clarity. As before, the first order one loop term in the vacuum wave functional follows, and is
\begin{equation}
  \includegraphics[height=2.9cm]{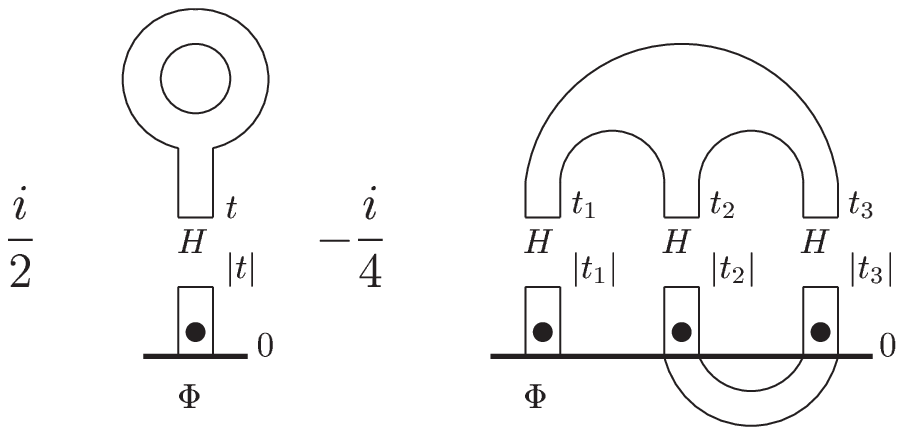}.
\end{equation}
This construction applies to both open and closed strings.

\section{The Schr\"odinger representation}

In the Schr\"odinger representation $n$--field correlation functions are expressed by $n$ functional integrals,
\begin{equation}\begin{split}\label{corr}
  \bra{0}\phi(\x_n,t_n)\cdots\phi(\x_1,t_1)\ket{0} = \int\pathD&(\phi_n\ldots\phi_1)\Psi_0[\phi_n]\phi_n(\x_n)\mathscr{S}[\phi_n,\phi_{n-1};t_n-t_{n-1}]\phi_{n-1} \\
  &\cdots \phi_2(\x_2)\mathscr{S}[\phi_2,\phi_1;t_2-t_1]\phi_{1}(\x_1)\Psi_0[\phi_1],
\end{split}\end{equation}
where there is one instance of the Schr\"odinger functional $\mathscr{S}$ for each product of two fields, and two instances of the vacuum state wave functional $\Psi_0$ at the initial and final times (though the time dependence is trivial since the vacuum is an eigenstate. The Schr\"odinger functional is defined by
\begin{equation}
  \mathscr{S}[\phi_2,\phi_1;t_2-t_1] = \bra{\phi_2}\exp\bigg(-\frac{i}{\hbar}\int\limits_{t_1}^{t_2}\!\ud x^0\,\, H(x^0)\bigg)\ket{\phi_1}.
\end{equation}
The Feynman description is
\begin{equation}
  \mathscr{S}[\phi_2,\phi_1;t] = \int\pathD\f\,\, \exp\bigg(\frac{i}{\hbar}\int\limits_{0}^t\!\ud x^0\,\, L(\varphi,\dot\f)\bigg)\bigg|_{\f=\phi_1\text{ at }x^0=0,}^{\f=\phi_2\text{ at }x^0=t}
\end{equation}
As before a shift of integration variable moves the field dependence out of the boundary conditions and into the action, giving
\begin{equation}\begin{split}
  \mathscr{S}[\phi_2,\phi_1;t] = \int\pathD\f\,\, \exp\bigg(\frac{i}{\hbar}\int\limits_{0}^t\!\ud x^0\,\, L(\varphi,\dot\f) &+ \frac{i}{\hbar}\int\!\ud^D\x\,\,\dot\f(\x,t)\phi_2(\x) \\
  &-\frac{i}{\hbar}\int\!\ud^D\x\,\,\dot{\f}(\x,0)\phi_1(\x)\bigg)\bigg|_{\f=0\text{ at }x^0=0}^{\f=0\text{ at }x^0=t}
\end{split}\end{equation}
The logarithm of the Schr\"odinger functional is the sum of connected Feynman diagrams constructed from a propagator $G_D$ which obeys Dirichlet boundary conditions at times $x^0=0$ and $x^0=t$ and with vertices integrated over the interval $x^0\in [0,t]$. All external legs end on the boundaries with a time derivative. The free field Schr\"odinger and vacuum functionals are therefore

\begin{equation}\label{sf-tree}
  \includegraphics[height=1.4cm]{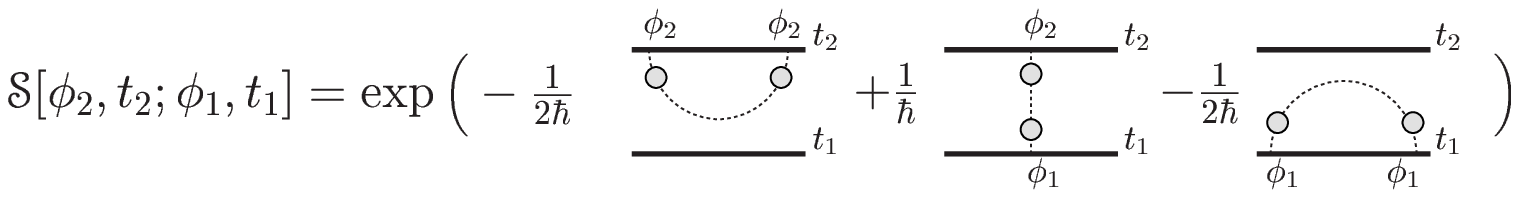},
\end{equation}
\begin{equation}\label{vwf-tree}
  \includegraphics[height=1.3cm]{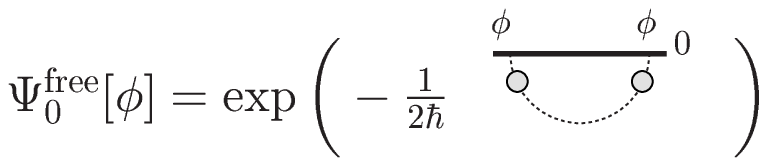},
\end{equation}
where a propagator drawn with a dotted line obeys Dirichlet conditions on all boundaries (heavy lines) shown in the diagram. The gray dot denotes a time derivative, and an unbroken line will represent the free space propagator. The method of images allows us to write the propagators in these diagrams as sums over free space propagators. For example, the propagator in the vacuum is
\begin{equation*}
  G_d(\x_2,t_2;\x_1,t_1) = G_0(\x_2,t_2;\x_1,t_1) - G_0(\x_2,t_2;\x_1,-t_1),
\end{equation*}
in terms of the free space propagator $G_0$. Using this it is a simple matter to show the equality of \rf{vwf-tree} and \rf{2.4}.

The equivalence of Schr\"odinger and Heisenberg representations of quantum theories is well documented, but less well known are the details of this equivalence. In this section we use our diagrammatic methods to demonstrate how covariant correlation functions are given by the non-covariant Schr\"odinger representation. We return to the standard local $\phi^3$ interaction and calculate the simplest non-trivial correlation function
\begin{equation}\label{3ptnorm}
  \includegraphics[width=0.85\textwidth]{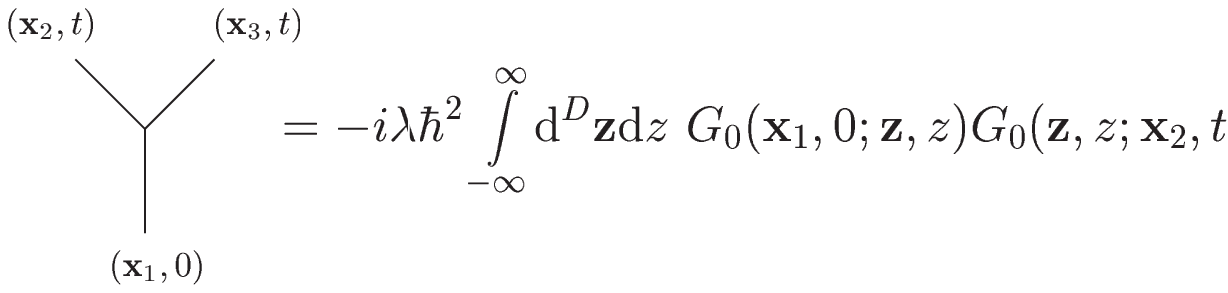}.
\end{equation}
In the Schr\"odinger representation this is given by the functional integral
\begin{equation}\label{3ptint}
 \int\pathD(\f_2,\f_1)\,\,\Psi_0[\f_2]\,\f_2(\x_3)\f_2(\x_2)\,\mathscr{S}[\f_2,\f_1;t]\,\f_1(\x_1)\Psi_0[\f_1].
\end{equation}
We need the first order Schr\"odinger and vacuum functionals, which are

\begin{equation}\label{s-pic}
 \includegraphics[width=0.75\textwidth]{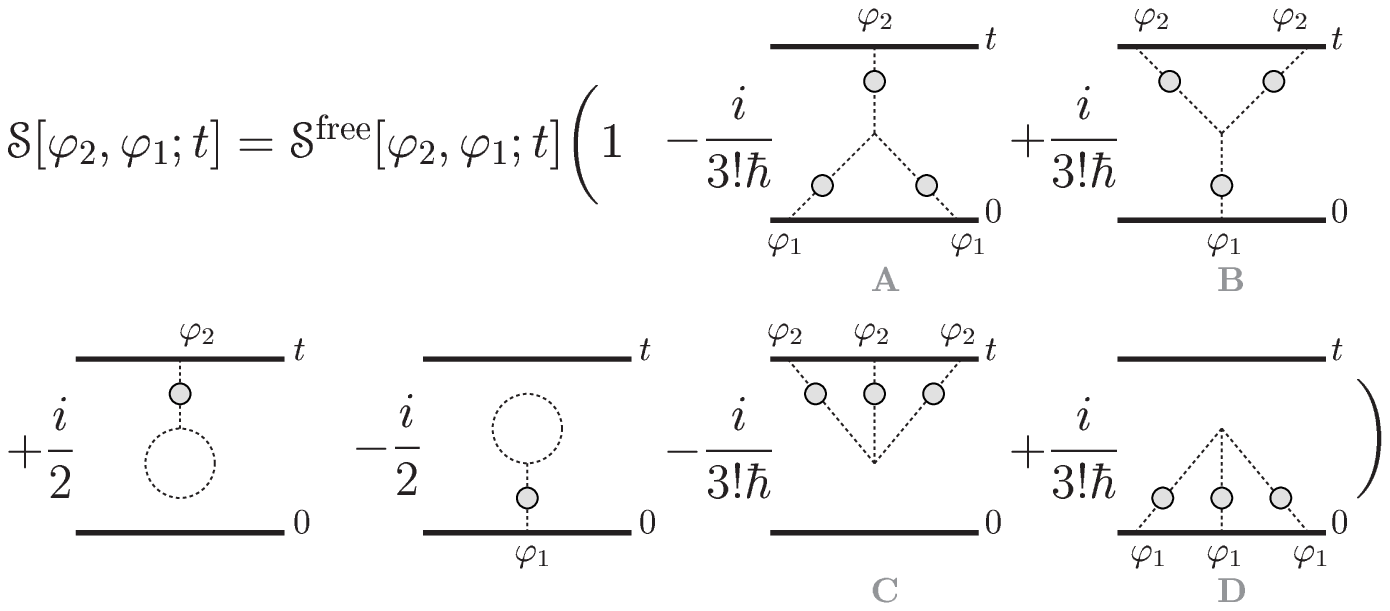}
\end{equation}

\begin{equation}\label{v-pic}
 \includegraphics[width=0.65\textwidth]{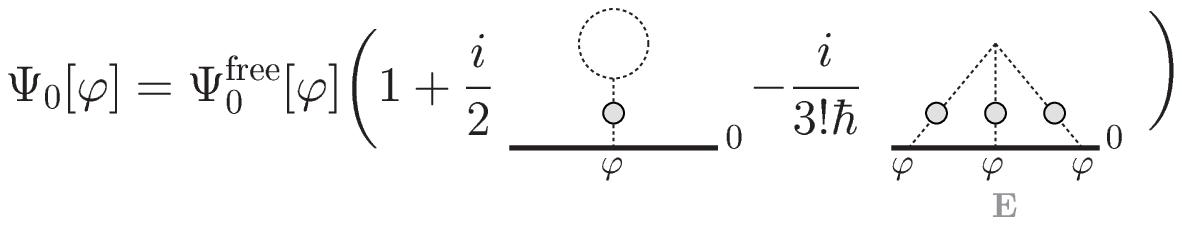}.
\end{equation}
The gray letters will be used to reference the diagrams. Excluding the loop diagrams, which can only contribute disconnected graphs, we must include all of the diagrams in (\ref{s-pic}) and (\ref{v-pic}) to calculate (\ref{3ptnorm}), not just diagram B which pictorially matches the desired result\footnote{The reason is that, besides being of the correct order to contribute, the boundary conditions on the Dirichlet propagators mean that even though they may not end on one of the boundaries they can still see it, and can contribute to the free space result}.

Inserting expressions (\ref{s-pic}) and (\ref{v-pic}) into (\ref{3ptint}) and carrying out the functional integrals the terms A to E contribute the set of diagrams shown below (excluding loops or disconnected pieces generated by the integrations). To illustrate we will give the calculation for diagram C explicitly, and state the results for the remaining diagrams. The numbers $1$ to $3$ will indicate the spatial positions of the external legs. The thick gray line is shorthand for $K^{-1}$, the inverse of the quadratic piece of the product of the Schr\"odinger functional and vacuum wave functional which contracts indices in the first functional integral of (\ref{3ptint}),

\begin{equation*}
  \includegraphics[height=5.5cm]{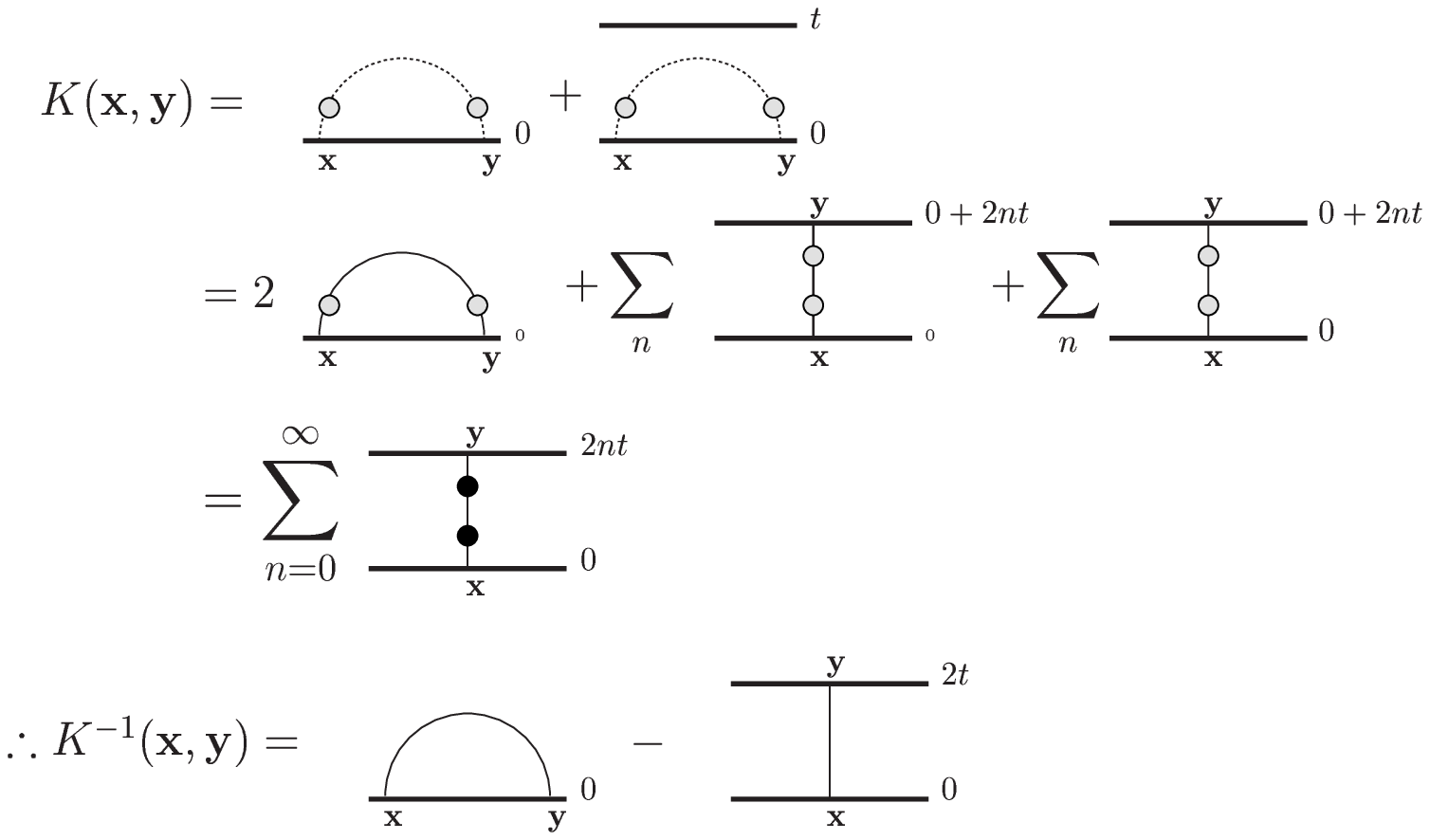}
\end{equation*}
We now begin the computation of diagram C. From (\ref{3ptint}) the $\varphi_1$ integral does not see the contribution from the interacting Schr\"odinger functional, and is Gaussian with the insertion $\varphi_1(\x_1)$,

\begin{equation*}
  \includegraphics[height=3.6cm]{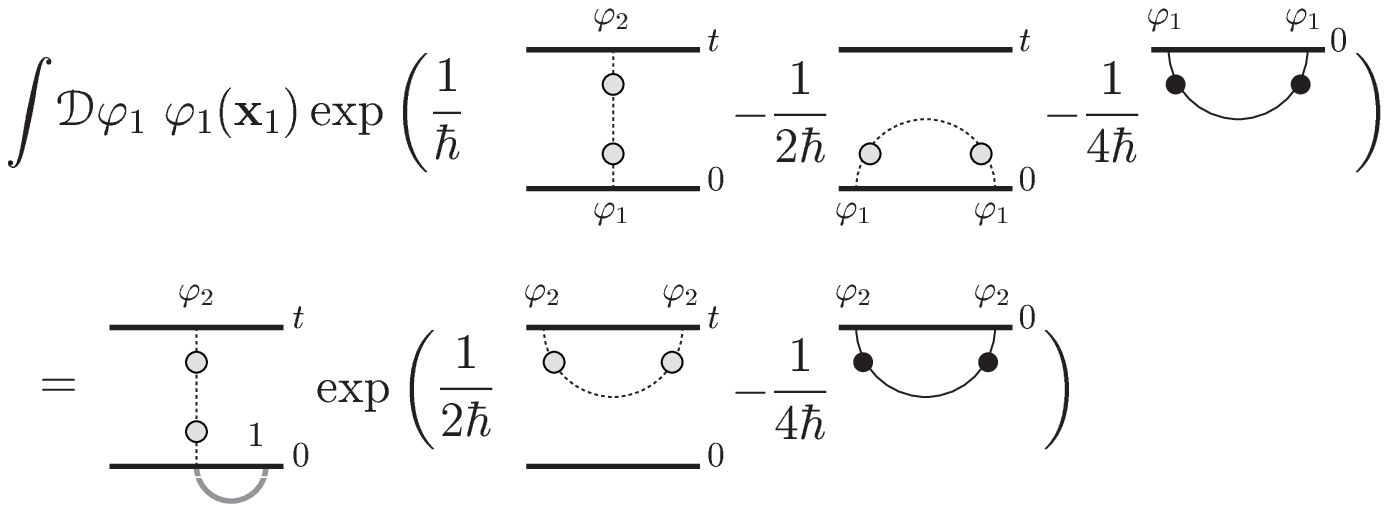}
\end{equation*}

The $\varphi_2$ integral becomes

\begin{equation*}
  \includegraphics[height=1.8cm]{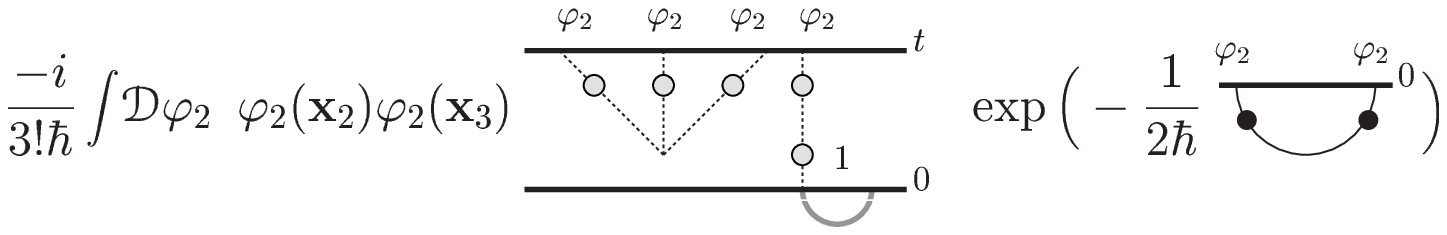}
\end{equation*}
To obtain a connected graph we must contract $\varphi_2$, $\varphi_3$ with the 3 point function and the remaining 3 point function field must be contracted with the field attached to $\x_1$. There are $3!$ equivalent ways of doing this, so the result is

\begin{equation*}
 \includegraphics[height=2.5cm]{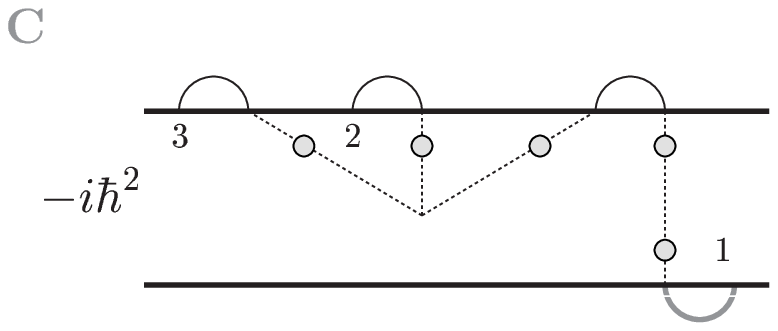}.
\end{equation*}
Using the method of images the propagator $G_D$ attached to one boundary is

\begin{equation}
  \includegraphics[height=1.9cm]{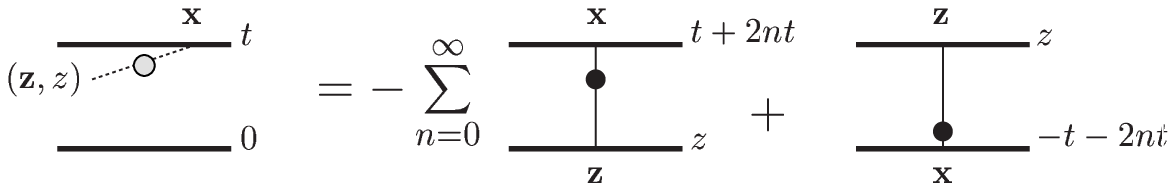}
\end{equation}
From which the gluing properties tell us that the two types of terms we encounter in diagram C are

\begin{equation}
  \includegraphics[height=1.8cm]{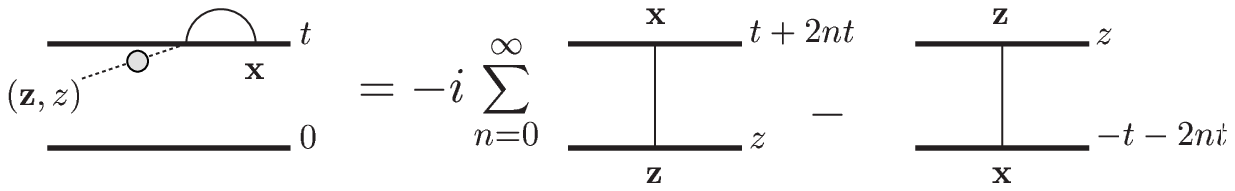}
\end{equation}
and

\begin{equation}
  \includegraphics[height=2cm]{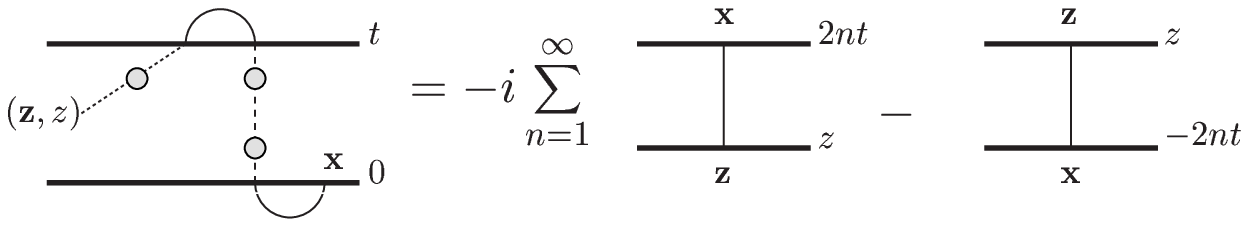}.
\end{equation}

Defining the following sets of diagrams,

\begin{equation}
 \includegraphics[height=1.4cm]{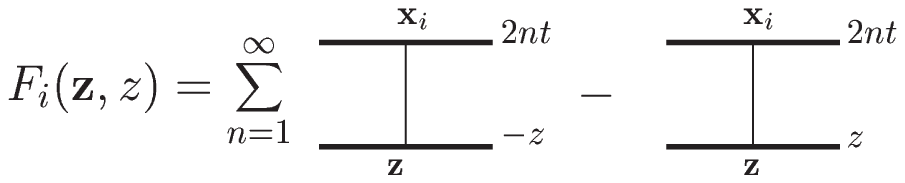},
\end{equation}

\begin{equation}
 \hspace{8pt}\includegraphics[height=1.4cm]{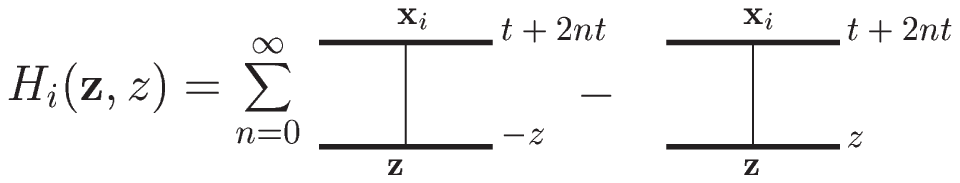}.
\end{equation}
Diagram C can be written
\begin{equation*}
  i\lambda\hbar^2 \!\!\int\limits_0^t \!\ud z \!\! \int\!\ud^D\z\,\, F_1(\z,z)H_2(\z,z)H_3(\z,z).
\end{equation*}
The calculation of the remaining diagrams A to E are similar and the results are below. There is no content in the length of the $K^{-1}$ lines which varies in the diagrams only for clarity. $K^{-1}$ will appear on the lower boundary and the equal time propagator on the upper boundary if we perform the $\varphi_1$ integral first, and vice versa if we perform the $\varphi_2$ integral first.

Diagram A contributes

\begin{equation*}
 \includegraphics[height=7.2cm]{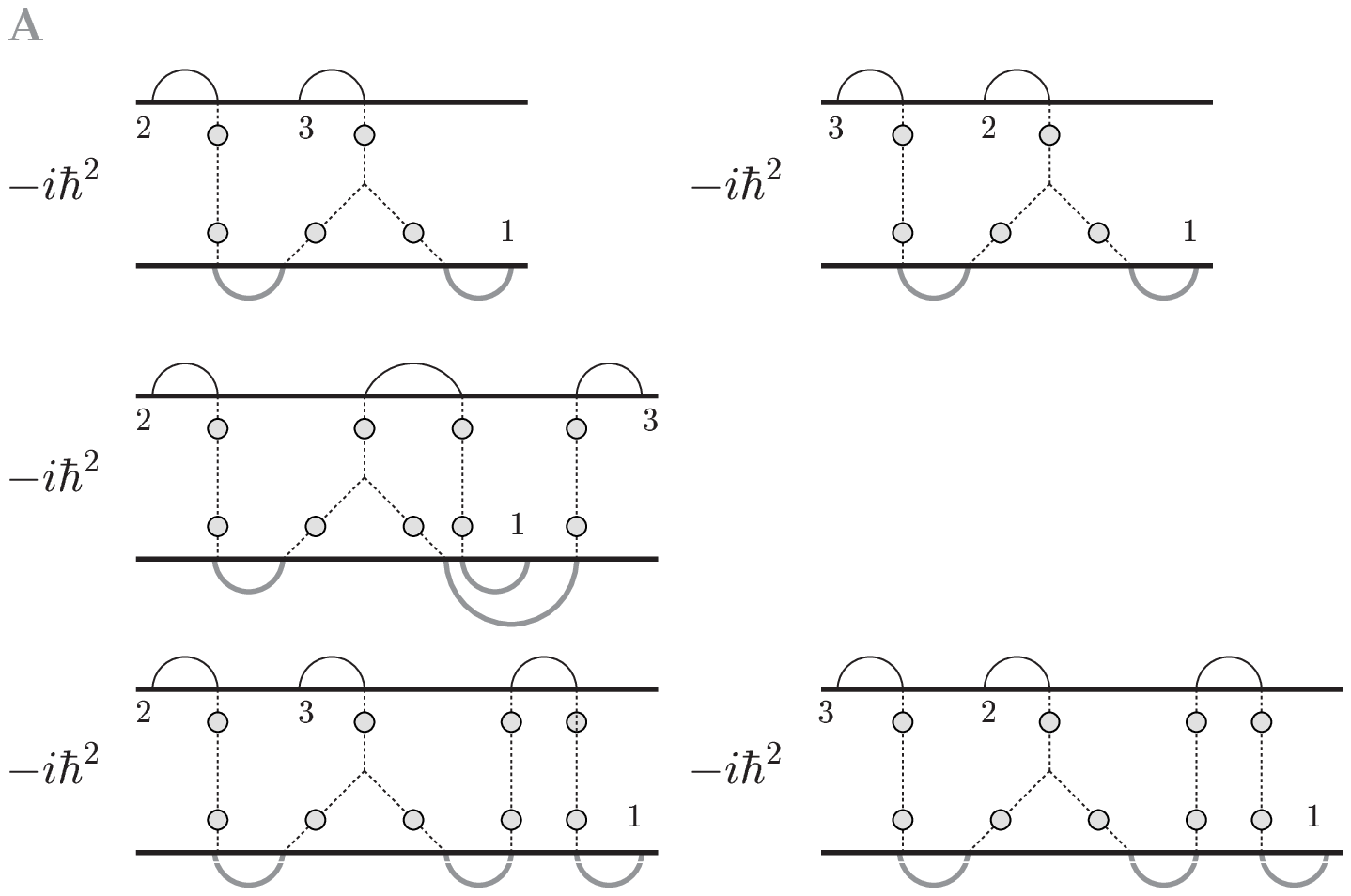}
\end{equation*}
Written in terms of the sums $F_i(\mathbf{z},z)$ and $H_i(\mathbf{z},z)$ this is (the order of the terms is respective to that of the diagrams above)
\begin{align*}
  i\lambda\hbar^2 \!\!\int\limits_0^t \!\ud z \!\! &\int\!\ud^D\z\big(G_0(\x_1,0;\z,z) - G_0(\x_1,2t;\z,z)\big) H_2(\z,z)\big( H_3(\z,z) + G_0(\z,t;\x_3,0)\big)\\
  &\hspace{8.85cm}+ (2\leftrightarrow 3)\\
  \\
  &+F_1(\z,z)\big(H_2(\z,z) + G_0(\z,z;\x_2,t)\big)\big( H_3(\z,z) + G_0(\z,z;\x_3,t)\big) \\
  \\
  &+\big(F_1(\z,z) + G_0(\x_1,2t;\z,z)\big)\big( H_2(\z,z) + G_0(\z,z;\x_2,t)\big)H_3(\z,z) + (2\leftrightarrow 3).
\end{align*}
Diagram B contributes

\begin{equation*}
 \includegraphics[height=5.5cm]{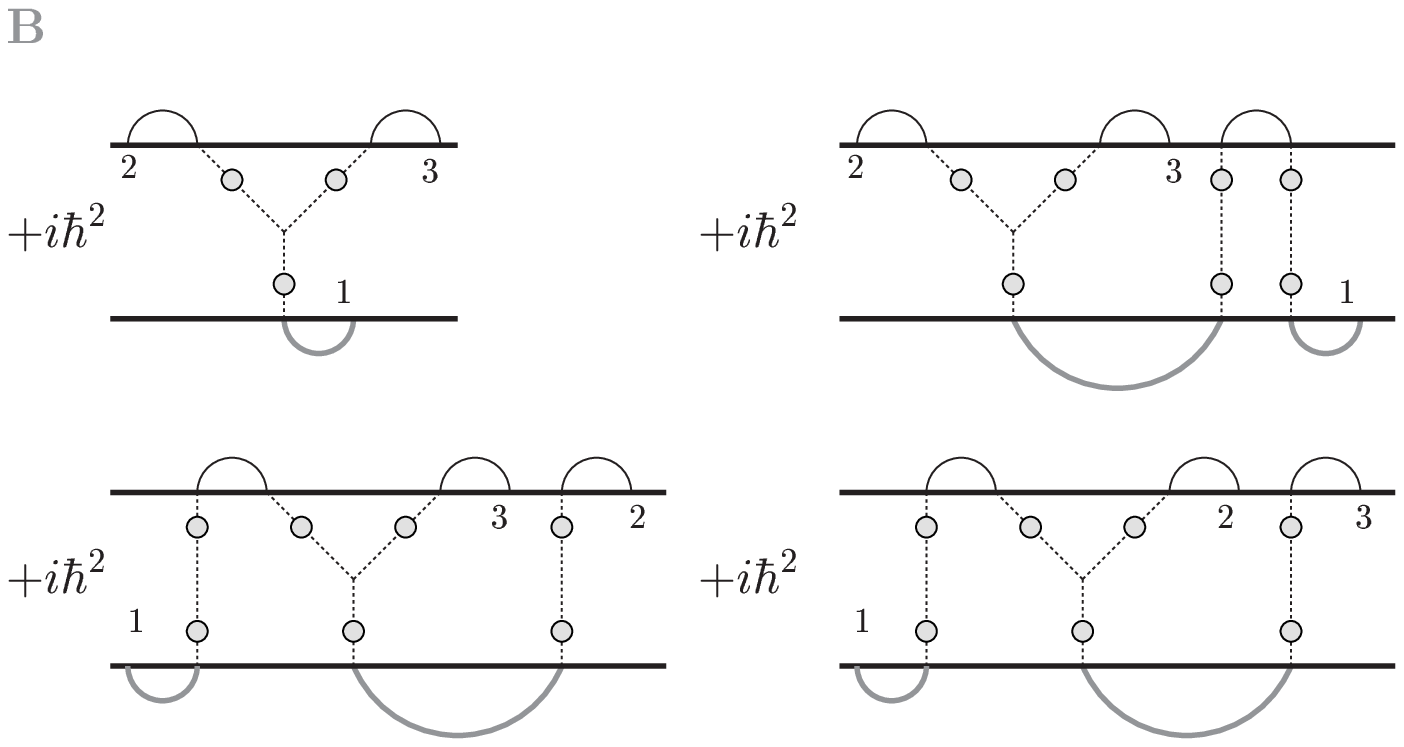}
\end{equation*}
which can be written
\begin{align*}
  -i\lambda\hbar^2 \!\!\int\limits_0^t \!\ud z \!\! &\int\!\ud^D\z\,\, \big(G_0(\x_1,0;\z,z) - G_0(\x_1,2t;\z,z)\big) H_2(\z,z)H_3(\z,z)\\
  &+ \big(F_1(\z,z) + G_0(\x_1,2t;\z,z)\big)H_2(\z,z)H_3(\z,z) \\
  &+ F_1(\z,z)\big(H_2(\z,z) + G_0(\z,z;\x_2,t\big)\big(H_2(\z,z) + G_0(\z,z;\x_2,t\big) H_3(\z,z) \\
  &\hspace{6.9cm}+ (2\leftrightarrow 3).
\end{align*}
Diagram D gives
\begin{equation*}
 \includegraphics[height=2.7cm]{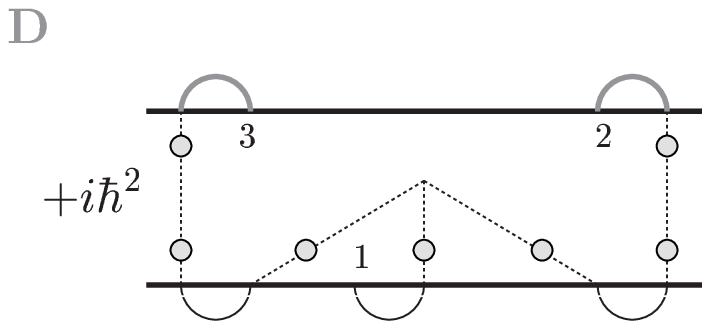}
\end{equation*}
which is 
\begin{equation*}\begin{split}
  -i\lambda\hbar^2 \!\!\int\limits_0^t \!\ud z \!\! \int\!\ud^D\z\,\, \big(F_1(\z,z) + G_0(\x_1,0;\z,z)\big) &\big(H_2(\z,z) + G_0(\z,z;\x_2,t)\big) \\
  &\times\big(H_3(\z,z) + G_0(\z,z;\x_3,t)\big).
\end{split}\end{equation*}
Before calculating the diagrams coming from the interacting vacuum functional it is worthwhile adding the above terms up to find, with no calculation other than cancellations, that the total contribution from the interacting Schr\"odinger functional and the free vacuum is
\begin{equation}\label{s-cont}
  -i\lambda\hbar^2 \!\!\int\limits_0^t \!\ud z \!\!\int\!\ud^D\z\,\, G_0(\x_1,0;\z,z)G_0(\z,z;\x_2,t)G_0(\z,z;\x_3,t).
\end{equation}
This is the correlation function we were looking for (with the correct coefficient) but the vertex is integrated in time only over the interval $[0,t]$. What remains must come from the vacuum. The contribution from the interacting vacuum functional $\Psi_0[\varphi_1]$, diagram E, is

\begin{equation*}
 \includegraphics[height=2.5cm]{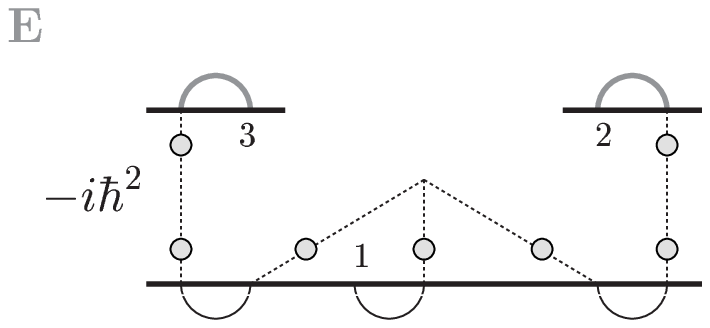}.
\end{equation*}
The upper boundary (at time $t$) is broken as a reminder that only the outermost Dirichlet propagators vanish on both boundaries (i.e. come from the free field Schr\"odinger functional), whereas the inner three propagators vanish only at $x^0=0$ (come from the free field vacuum). This is
\begin{equation}
  -i\lambda\hbar^2 \!\!\int\limits_{-\infty}^0 \!\ud z \!\!\int\!\ud^D\z\,\, G_0(\x_1,0;\z,z)G_0(\z,z;\x_2,t)G_0(\z,z;\x_3,t),
\end{equation}

There is a contribution from the vacuum $\Psi_0[\varphi_2]$, which we call term F,

\begin{equation*}
 \includegraphics[height=2.5cm]{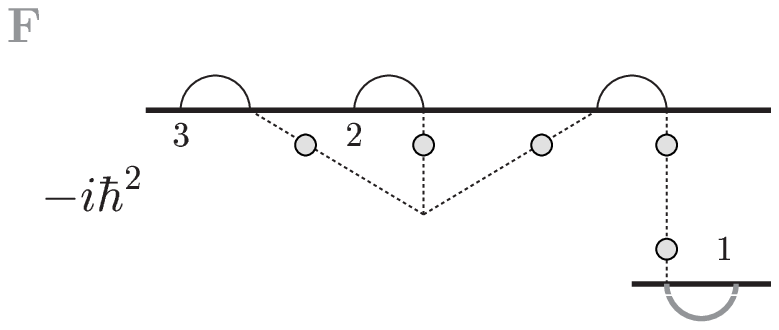}
\end{equation*}
Again, only the rightmost propagator comes from the Schr\"odinger functional. This is
\begin{equation}
  -i\lambda\hbar^2 \!\!\int\limits_{-\infty}^0 \!\ud z \!\!\int\!\ud^D\z\,\, G_0(\x_1,z;\z,t)G_0(\z,z;\x_2,0)G_0(\z,z;\x_3,0).
\end{equation}
Changing variable
\begin{equation}\label{cov}
  z\rightarrow -z+t
\end{equation}
in $F$ we have a total contribution of
\begin{equation}\begin{split}
-i\lambda\hbar^2 \!\!\int\limits_{-\infty}^0 \!\ud z \!\!\int\!\ud^D\z\,\, &G_0(\x_1,0;\z,z)G_0(\z,z;\x_2,t)G_0(\z,z;\x_3,t) \\
  &-i\lambda\hbar^2 \!\!\int\limits_t^\infty \!\ud z \!\!\int\!\ud^D\z\,\, G_0(\x_1,0;\z,z)G_0(\z,z;\x_2,t)G_0(\z,z;\x_3,t)
\end{split}\end{equation}
This is again the integrand we want, but we are missing the integral over the region $[0,t]$. This missing piece is precisely what we found to be contributed by the Schr\"odinger functional in (\ref{s-cont}).

We have seen that for this simple scattering process the interacting vacuum wave functional gives almost the correct result, up to some `small' missing piece (in the sense that term is integrated over only a finite time), and that this is given by the Schr\"odinger functional.

\subsection*{Conclusions}
We have described, in analogy with quantum field theory, how to construct the string field theory vacuum wave functional for both open and closed strings. Note that we have not given, nor do we need to give, the explicit form of the three string vertex. That choice is arbitrary, the required kernels could even, theoretically, be computed with the Polyakov integral on the relevant manifold. For example, for the open string kernel $\Gamma$ introduced in (\ref{Gamma}) this would be a disk with marked sections on the boundary. Although we have only given a few terms in the expansion of the vacuum functional thought we expect the construction to generalise to all orders in perturbation theory.

Given the gluing property for string fields we may postulate a string field theory interaction which is non-local in the spatial co-ordinates, but local in time, and where the fields depend only on $t$ and the set $\X$ defined in the introduction. For example we could define an interaction very like that in the light-cone gauge \cite{Kaku}, with worldsheet $\tau$ replaced by spacetime $t$,
\begin{equation}\begin{split}
  \int\limits_{-\infty}^\infty\!\ud t\!\!\int\pathD(\X, \X_1, \X_2)\,\,\Phi[&\X(\sigma),t]\Phi[\X_1(\sigma_1),t]\Phi[\X_2(\sigma_2),t] \\
  &\prod\limits_{0\leq\sigma\leq\pi/2}\delta(\X(\sigma) - \X_1(\sigma_1))\prod\limits_{\pi/2\leq\sigma\leq\pi}\delta(\X(\sigma) - \X_2(\sigma_2))
\end{split}\end{equation}
where all three fields live at the same time (as in a particle theory) and the parameters along the string are
\begin{equation}\begin{split}
  \sigma_1 &= 2\sigma,\quad 0\leq\sigma\leq\pi/2 \\
  \sigma_2 &= 2\sigma-\pi,\quad \pi/2\leq\sigma\leq\pi
\end{split}\end{equation}
so that each string has parameter domain $0\leq\sigma\leq\pi$. If the free Hamiltonian was the inverse of $G(\X_2,t_2;\X_1,t_1)$ we could repeat any quantum field theory argument in this theory. However, this approach removes the $X^0$ oscillators from the theory at the outset, and it is unclear how to justify this nor how amplitudes in this theory relate to known results.

The interaction is local in time, but as with the light-cone gauge we do not expect locality in one direction to spoil the UV finiteness of string interactions \cite{Eliezer}. Our time co-ordinate, unlike in other string field theories, is a time at which the whole spatially extended string (with ghosts) exists. This seems to work around the problems \cite{Maeno} of, for example, quantising Witten's theory \cite{Witten} where time is normally taken to be the midpoint of $X^0(\sigma)$, but the string remains extended in time. This may be a worthwhile avenue for future study.

Although it is a lengthier task, it is not especially more difficult to reconstruct the Schr\"odinger functional of quantum field theory using a similar approach, and the generalisation to string field theory will follow. We would propose the first order expansion 
\begin{equation}
 \mathscr{S}[\phi_2,\phi_1,t]=\mathscr{S}^\text{free}[\phi_2,\phi_1,t]\bigg(1+\phi_2\phi_2\phi_2\mathscr{S}^{(3,0)} + \phi_2\phi_2\mathscr{S}^{(2,1)}\phi_1 + \phi_2\mathscr{S}^{(1,2)}\phi_1\phi_1 + \mathscr{S}^{(0,3)}\phi_1\phi_1\phi_1\bigg)
\end{equation}
which includes four unknown kernels, $\mathscr{S}^{(i,j)}$. There are four possible three field amplitudes which must be reproduced by a functional integration with one instance the Schr\"odinger functional,
\begin{align*}
  &\bra{0}\phi(\x,t)\phi(\y,t)\phi(\z,t)\ket{0}& &\bra{0}\phi(\x,t)\phi(\y,t)\phi(\z,0)\ket{0}& \\
  &\bra{0}\phi(\x,t)\phi(\y,0)\phi(\z,0)\ket{0}& &\bra{0}\phi(\x,0)\phi(\y,0)\phi(\z,0)\ket{0}&
\end{align*}
giving us four simultaneous functional equations with which to determine the kernels $\mathscr{S}^{(i,j)}$.

\end{document}